\documentclass[runningheads]{llncs}
\usepackage{graphicx}
\usepackage{xcolor}
\usepackage{hyperref}
\usepackage{makecell}
\usepackage{caption}
\usepackage{subcaption}
\newcommand{\pcdiff}[1]{\textcolor{red}{\emph{{#1}}}}

\begin{document}
\title{Performance characterisation of the 64-core SG2042 RISC-V CPU for HPC}
%
%
\author{Nick Brown\inst{1}\orcidID{0000-0003-2925-7275} \and Maurice Jamieson\inst{1}\orcidID{0000-0003-1626-4871}}
\authorrunning{N. Brown}
%
\institute{EPCC at the University of Edinburgh, 47 Potterrow, Edinburgh, UK }
\maketitle              
\begin{abstract}
Whilst RISC-V has grown phenomenally quickly in embedded computing, it is yet to gain significant traction in High Performance Computing (HPC). However, as we move further into the exascale era, the flexibility offered by RISC-V has the potential to be very beneficial in future supercomputers especially as the community places an increased emphasis on decarbonising its workloads. Sophon's SG2042 is the first mass produced, commodity available, high-core count RISC-V CPU designed for high performance workloads. First released in summer 2023, and at the time of writing now becoming widely available, a key question is whether this is a realistic proposition for HPC applications.

In this paper we use NASA's NAS Parallel Benchmark (NPB) suite to characterise performance of the SG2042 against other CPUs implementing the RISC-V, x86-64, and AArch64 ISAs. We find that the SG2042 consistently outperforms all other RISC-V solutions, delivering between a 2.6 and 16.7 performance improvement at the single core level. When compared against the x86-64 and AArch64 CPUs, which are commonplace for high performance workloads, we find that the SG2042 performs comparatively well with computationally bound algorithms but decreases in relative performance when the algorithms are memory bandwidth or latency bound. Based on this work, we identify that performance of the SG2042's memory subsystem is the greatest bottleneck.

\end{abstract}
\section{Introduction}
RISC-V is an open Instruction Set Architecture (ISA) that, since it was first released over a decade ago, has gained significant traction. At the time of writing it was recently announced that over 13 billion RISC-V CPU cores have been manufactured, but many of these are in embedded computing such as automotive, space, and micro-controllers. RISC-V has yet to become commonplace in High Performance Computing (HPC), but as the HPC community moves further into the exascale era and there is an increased emphasis on decarbonisation of workloads, we need to consider how to best deliver both increased performance and greater energy efficiency. To this end, there is a renewed interest in new hardware solutions and technologies built atop RISC-V have a strong potential here as they can offer specialisation whilst still providing a common software ecosystem.

Sophon's SG2042 is the first high core count commodity available RISC-V CPU designed for high performance workloads. First released in summer 2023, this mass produced, 64-core RISC-V CPU is aimed at high performance workloads. Not only does this processor provide significantly more cores that existing, SoC based, mass produced RISC-V CPUs, but furthermore the T-Head XuanTie C920 cores themselves have been designed for high performance. Consequently this new RISC-V CPU is very interesting to the HPC community and previous work \cite{brown2023risc} found that, for the RAJAPerf suite \cite{hornung2017raja}, it delivers a considerable performance uplift compared to existing commodity available RISC-V CPUs, but struggled to match a set of x86-based CPUs that are commonplace in HPC machines. In this paper we leverage NASA's NAS Parallel Benchmark (NPB) suite to undertake more in depth performance characterisation of the SG2042. Running this suite across CPUs that implement the RISC-V, x86-86 and AArch64 ISAs, and in the later two categories because we have selected CPUs that are used in production supercomputers, we are able to better understand the types of workloads that the SG2042 suits and where it might fall short.

\section{Background}
\subsection{The Sophon SG2042}
The Sophon SG2042 CPU is a 64-core processor running at 2GHz and organised in clusters of four XuanTie C920 cores. Each 64-bit core, designed by T-Head, is designed for high performance workloads and adopts a 12-stage out-of-order multiple issue superscalar pipeline design \cite{c906}. Implementing the RV64GCV instruction set, the C920 has three decode, four rename/dispatch, eight issue/execute and two load/store execution units. Version 0.7.1 of the vectorisation standard extension (RVV v0.7.1) is supported \cite{c920-nofp64}, with a vector width of 128 bits. Each C920 core contains 64KB of L1 instruction (I) and data (D) cache, 1MB of L2 cache which is shared between the cluster of four cores, and 64MB of L3 system cache which is shared by all cores in the package. The SG2042 also provides four DDR4-3200 memory controllers, and 32 lanes of PCI-E Gen4. The CPU we use for the benchmarking in this paper is contained in a Pioneer Box by Milk-V which has 128GB of DDR4 RAM. 

The SG2042's C920 core only provides RVV v0.7.1 which is not supported by mainline GCC or LLVM. To this end, T-Head have provided their own fork of the GNU compiler (XuanTie GCC) which has been optimised for their processors and supports RVV v0.7.1. It has been found \cite{lee2023backporting} that GCC8.4, which is part of their 20210618 release, provides the best auto-vectorisation capability and-so this is the version we use for the benchmarking experiments undertaken in this paper. Their version of the compiler generates Vector Length Specific (VLS) RVV assembly which specifically targets the 128-bit vector width of the C920. All codes are compiled at optimisation level three, and all reported results are averaged over five runs. At the time of execution each benchmark run reported in this paper was making exclusive use of the machine.

In \cite{brown2023risc} the authors benchmarked the SG2042 using the RAJAPerf suite, however this was across a large number of individual kernels and from the results it was difficult to isolate and identify individual performance patterns. By contrast, in this paper we characterise and explore each individual benchmark of the NPB suite to better classify the performance properties of the SG2042.

\subsection{NAS Parallel Benchmarks (NPB) suite}
\label{sec:npb}
The NAS Parallel Benchmark (NPB) suite \cite{benchmarks2006parallel} is a collection of benchmarks developed by NASA's Advanced Supercomputing (NAS) division to characterise HPC systems, especially for Computational Fluid Dynamics (CFD) applications. First released in the mid 1990s, in this paper we leverage the original eight benchmarks in the suite, which comprises five kernels and three pseudo applications. The kernels capture key algorithmic patterns that are ubiquitous throughout HPC codes and test key performance characteristics that are important across many workloads. The pseudo applications combine multiple kernels to provide more complicated workloads. All these benchmarks are configured using a variety of problem sizes known as classes. There are a variety of implementations of the suite provided by NAS, including the OpenMP and MPI versions that we use here, and throughout this paper use the official code without any modifications.

\begin{table}[]
    \centering
    \caption{Summary of memory behaviour for NPB benchmarks on a Xeon Platinum 8170}
    \label{tab:npb_description}
    \begin{tabular}{|c|c|c|c|}
    \hline
      \textbf{Benchmark}   & \makecell{\textbf{Clock ticks} \\ \textbf{cache stall}} & \makecell{\textbf{Clock ticks} \\ \textbf{DDR stall}} & \makecell{\textbf{Time DDR} \\ \textbf{bandwidth bound}} \\
     \hline
	  \textbf{Integer Sort (IS)} & 35\% & 0\% & 16\% \\        
	  \textbf{Multi Grid (MG)} & 34\% & 20\% & 88\% \\
        \textbf{Embarrassingly Parallel (EP)} & 11\% & 0\% & 0\% \\
        \textbf{Conjugate Gradient (CG)} & 19\% & 18\% & 0\% \\
        \textbf{Fast Fourier Transform (FT)} & 13\% & 9\% & 18\% \\
        \hline
        \textbf{Block Tridiagonal (BT)} & 8\% & 9\% & 0\% \\
        \textbf{LU Gauss Seidel (LU)} & 12\% & 11\% & 0\% \\
        \textbf{Scalar Pentadiagonal (SP)} & 20\% & 21\% & 0\% \\
        
    \hline
    \end{tabular}
\end{table}

Table \ref{tab:npb_description} summarises, for each benchmark in the suite, the memory behaviour when run using OpenMP on all 26 physical cores of a Xeon Platinum 8170. The \emph{Clock ticks cache stall} and \emph{Clock ticks DDR stall} columns report how often the CPU was stalled on cache and main memory accesses respectively, and the \emph{Time DDR bandwidth bound} column reports the percentage of execution time that there was a high DDR bandwidth utilisation.

The IS kernel tests indirect, random, memory accesses which it can be seen stalls a significant fraction of the CPU due to cache accesses. It can be seen that the MG kernel is heavily memory bound both in terms of time stalled on cache and main memory accesses, and also the percentage of execution time where DDR is under high utilisation. By contrast, the EP benchmark is designed to test compute performance and there are far fewer cycles stalled on memory access, and no time spent with high DDR bandwidth utilisation. CG comprises irregular memory access and nearest neighbour communication, which results in around 37\% of clock ticks stalled on cache or DDR accesses, and the FT benchmark requires all-to-all communications between ranks to undertake a parallel transposition of data. For FT it can be seen that whilst there is only 22\% of clock ticks stalled, which is lower than the five other kernels apart from EP, the kernel is utilising a high DDR bandwidth for 18\% of the time.

The BT, LU and SP pseudo application benchmarks are more complicated than the five NPB kernels, and represent common, real-world, HPC use-cases. All three of these pseudo applications compute a finite difference solution to the 3D compressible Navier Stokes equations, where the LU benchmark solves this via a block-lower block-upper triangular approximation based upon Gauss Seidel iterative method \cite{saphir1997new}. The BT and SP benchmarks solve the same problem as LU, but base their solution on a Beam-Warming approximation. In BT the resulting equations are block-tridiagonal whereas in SP are fully diagonalised \cite{saphir1997new}. Both these systems are solved using Gaussian elimination. It can be seen from Table \ref{tab:npb_description} that, out of these three pseudo applications, BT stalls the least on memory accesses and SP the most. 

\section{RISC-V core comparison}
\label{sec:riscv-comparison}

In this section we compare performance of existing commodity RISC-V solutions. Due to the difference in core counts between RISC-V CPUs, we focus here on single cores performance to understand how the XuanTie C920 core of the Sophon SG2042 performs against other widely available RISC-V cores. We compare against the U74 core \cite{u74} which is contained in the JH7200 and JH7100 SoCs of the VisionFive V2 and V1 respectively, and both of these boards contain 8GB of DRAM. We also compare against the SiFive Freedom U740 SoC, also containing the U74 core and 16GB of DDR, and the T-Head XuanTie C906 \cite{c906} in the AllWinner D1 SoC with 1GB of memory. 

\begin{table}[]
    \centering
    \caption{Single core comparison between RISC-V technologies with performance reported in Mops/s (Higher is better) using NPB kernels running at class B. In red is the percentage performance delivered compared to the C920 core of the SG2042.}
    \label{tab:risc-v-core-comparison}
    \begin{tabular}{|c|c|c|c|c|c|}
    \hline
      \textbf{Benchmark}   & \textbf{SG2042} & \textbf{VisionFive V2} & \textbf{VisionFive V1} & \textbf{SiFive U740} & \textbf{All Winner D1}\\
     \hline
	IS & 60.6 & \makecell{17.84 \\ \pcdiff{(29\%)}} & \makecell{6.36 \\ \pcdiff{(10\%)}} & \makecell{9.09 \\ \pcdiff{(15\%)}} & \makecell{5.41  \\ \pcdiff{(9\%)}}\\        
	MG & 1210.05 & \makecell{288.65 \\ \pcdiff{(24\%)}} & \makecell{72.31 \\ \pcdiff{(6\%)}} & \makecell{90.28 \\ \pcdiff{(7\%)}} & \makecell{163.19 \\ \pcdiff{(13\%)}}\\
    EP & 31.35 & \makecell{12.01 \\ \pcdiff{(38\%)}} & \makecell{7.55 \\ \pcdiff{(24\%)}} & \makecell{9.08 \\ \pcdiff{(29\%)}} & \makecell{9.23 \\ \pcdiff{(29\%)}}\\
        CG & 205.25 & \makecell{43.61 \\ \pcdiff{(21\%)}} & \makecell{21.96 \\ \pcdiff{(11\%)}} & \makecell{20.09 \\ \pcdiff{(10\%)}} & \makecell{12.99 \\ \pcdiff{(6\%)}} \\
        FT & 857.64 & \makecell{245.99 \\ \pcdiff{(29\%)}} & \makecell{88.35 \\ \pcdiff{(10\%)}} & \makecell{116.59 \\ \pcdiff{(14\%)}} & DNR \\
        
    \hline
    \end{tabular}
\end{table}

Table \ref{tab:risc-v-core-comparison} reports a single core performance comparison between these RISC-V technologies, for the five NPB kernels at class B, with performance reported in million operations per second (Mop/s) and a higher number is better. In italicised red is the percentage performance that a single core of this CPU delivers compared to a single C920 core found in the SG2042. It can be seen that, irrespective of the kernel, the C920 significantly out performs all other RISC-V technologies. Consistently, the U74 of the VisionFive V2 performs closest to the C920, but is still only delivering between 21\% and 38\% the performance of the C920. Whilst the VisionFive V1 and SiFive U740 both contain the same U74 core as the VisionFive V2, they are significantly slower and this is broadly in agreement with \cite{brown2023risc}. 

The C906 of the All Winner D1 is out performed by the C920 and the U74 of the V2 quite considerably. However, this is the cheapest of the SoCs considered here, and the C906 outperforms the V1 and U740 for the EP and MG benchmarks. Given the performance profile of the benchmarks reported in Table \ref{tab:npb_description}, this suggests that the raw compute power of the C906 is similar to that of the U74 and the memory bandwidth is greater on the All Winner D1 than the VisionFive V1 and SiFive U740. However, for those benchmarks with more complex, irregular, memory patterns such as IS and CG, the C906 seems to struggle compared to the other RISC-V cores. Incidentally, it was not possible to run the FT benchmark on the All Winner D1 due to the limited 1GB of memory becoming exhausted.

In this section we therefore conclude that the C920 core of the SG2042 significantly outperforms all other commodity available RISC-V CPU cores. Whilst this is in agreement with \cite{brown2023risc}, in this section we have compared against a wider range of RISC-V CPUs than \cite{brown2023risc} and for specific algorithmic patterns that are very commonly found in HPC codes, especially for CFD. Considering that they use the same U74 core, it is surprising that the VisionFive V2 outperforms the V1 and U740 by such a large margin, but again this is in agreement with \cite{brown2023risc} and \cite{lee2023test}, and one of the reasons for this is that the V2 is running at 1.5GHz compared to 1.2GHz for both the V1 and U740. 

\section{Comparing the SG2042 against other architectures}
\label{sec:otherarch}
In Section \ref{sec:riscv-comparison} we compared the performance of the SG2042's C920 core against other RISC-V commodity available CPU cores. Whilst it is interesting to explore performance against RISC-V CPUs, and indeed the C920 delivers impressive performance compared to other RISC-V hardware, to understand whether the SG2042 is a contender for HPC it is far more instructive to benchmark against CPUs of other architectures that are commonly used for HPC workloads. 

\begin{table}[]
    \centering
    \caption{Summary of CPUs that are benchmarked in this section}
    \label{tab:other_arch}
    \begin{tabular}{|c|c|c|c|c|c|}
    \hline
      \textbf{CPU} & \textbf{ISA} & \textbf{Part} & \textbf{Base clock} & \makecell{\textbf{Number} \\ \textbf{of cores}} & \textbf{Vector} \\
     \hline
	 AMD EPYC & x86-64 & EPYC 7742 & 2.25GHz & 64 & AVX2\\
      Intel Skylake & x86-64 & Xeon Platinum 8170 & 2.1 GHz & 26 & AVX512\\      
      Marvell ThunderX2 & ARMv8.1 & CN9980 & 2 GHz & 32 & NEON\\
      Sophon SG2042 & RV64GCV & SG2042 & 2 GHz & 64 & RVV v0.7.1\\
    \hline
    \end{tabular}
\end{table}

In this section we compare against CPUs of other architectures that are commonplace in HPC, and these are summarised in Table \ref{tab:other_arch}. The AMD EPYC is the Rome series of AMD CPUs, containing the Zen-2 micro architecture and we run this on ARCHER2, a Cray EX and the UK national supercomputer. Similarly to the SG2042, the AMD EPYC contains 64 physical cores across four NUMA regions, each with 16 cores, but has eight instead of four memory controllers and memory channels. Each core in the AMD EPYC contains 32KB of I and D L1 cache, 512 KB of L2 cache, and there is 16MB of L3 cache shared between four cores. Providing AVX2, the EPYC 7742 has 256-bit wide vector registers, which is double that of the SG2042, but is capable of processing two AVX-256 instructions per cycle. ARCHER2 contains 256GB of DDR memory. We use GCC version 11.2 when compiling on ARCHER2. Simultaneous Multithreading (SMT) is disabled for our runs, which is the default configuration on ARCHER2.

We also compare against an Intel Skylake Xeon Platinum 8170, which is the same CPU used to profile the NPB benchmarks in Table \ref{tab:npb_description}. This Skylake-SP CPU contains 26 cores, each with 32KB of I and D L1 cache, 1MB of L2 cache and 1.375MB of L3 cache (the later is shared across all cores). The Skylake supports AVX512, double and quadruple the width of the EPYC 7742 and SG2042 respectively, and each Skylake core has two FPUs. The machine we run on has 192GB of DDR4 memory, and we use GCC version 8.4.

Lastly, we compare against the CN9980 Marvell ThunderX2 which contains 32 cores implementing the ARMv8.1 (AArch64) ISA via the Vulcan micro architecture. Each core contains 32KB of I and D L1 cache, as well as 256KB of L2. There is a total of 32MB L3 cache, 1MB per core, shared by the entire chip. NEON is supported, which provides 128 bit wide vector registers and this is interesting because it matches the vector width of the C920 core in the SG2042. Similarly to the Skylake, the Marvell ThunderX2 has two FPUs per core. This is the CPU used in Fulhame, an HPE Apollo 70 system, with 128GB of DDR per node, we use GCC version 9.2 and SMT is also disabled in our runs. 

For the performance comparison undertaken in this section, we run class C of the NASA Parallel Benchmarks and run over multiple cores of the CPUs by using the OpenMP implementations of the benchmarks \cite{jin1999openmp}. Each thread is mapped to an individual physical CPU core, all reported results are averaged over five runs and all codes are built at optimisation level three.

\subsection{Integer Sort (IS)}

\begin{figure*}[htb]
    \centering
    {\includegraphics[width=0.8\textwidth]{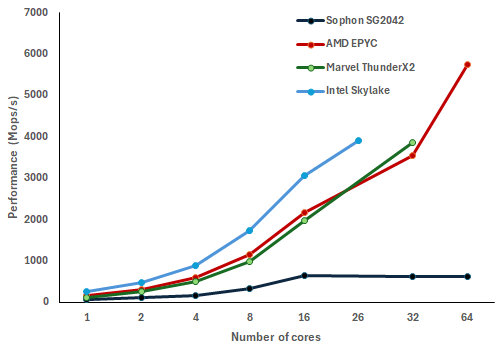}}
    \caption{IS benchmark performance (higher is better) parallelised via OpenMP}
    \label{fig:is}
\end{figure*}

As described in Section \ref{sec:npb}, the Integer Sort (IS) benchmark is concerned with integer comparison and indirect, random, memory access performance. Figure \ref{fig:is} illustrates the performance results for this benchmarks across our CPUs of interested, reported in Mops/s (higher is better). It can be seen that the SG2042 performs considerably worse than all other CPUs with performance plateauing at 16 cores. By contrast, the ThunderX2 and AMD EPYC delivering similar performance until the 32 cores of the ThunderX2 are exhausted. The Skylake performs better than all the other CPUs, but is limited by its lower core count, where the ThunderX2 catches up to the Skylake at 32 cores and the EPYC outperforms it at 64 cores. 

It can be seen from Figure \ref{fig:is} that the SG2042 struggles significantly with this benchmark where as was seen in Table \ref{tab:npb_description} the irregular, random, memory accesses result in a comparatively large number of time stalled due to cache access and DDR bandwidth utilisation is high for a small fraction of the runtime. A hypothesis is that could be due to the cache hierarchy, where the Skylake which performs the best has the largest L2 cache, 1MB per core, compared to 256KB (per core, 1MB shared between four cores) for the SG2042, 256KB for the ThunderX2 and 512KB for the AMD EPYC. The surprise here is in the performance difference between the SG2042 and the ThunderX2, as per core they both have the same amount of L2 and L3 cache. 

\subsection{Multi Grid (MG)}
\begin{figure*}[htb]
    \centering
    {\includegraphics[width=0.8\textwidth]{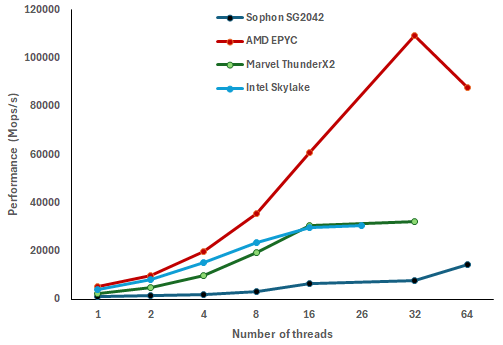}}
    \caption{MG benchmark performance (higher is better) parallelised via OpenMP}
    \label{fig:mg}
\end{figure*}

It was illustrated in Section \ref{sec:npb} that the Multi Grid (MG) benchmark is heavily memory bandwidth bound, and results of executing this benchmark kernel on the CPUs of interest is illustrated in Figure \ref{fig:mg}. It can be seen that the AMD EPYC provides considerably best performance, with the Skylake and ThunderX2 delivering similar performance and both plateauing at 16 cores where memory bandwidth is likely saturated. By contrast, the SG2042 lags the other CPUs considerably, also plateauing at 16 and 32 cores but then with a performance increase at 64 cores.

The memory configuration of the CPUs partially helps to explain the relative performance reported in Figure \ref{fig:mg}. The AMD EPYC has 8 memory controllers and 8 memory channels, connected to DDR4-3200 memory. By contrast, the Skylake and ThunderX2 both only have 2 memory controllers and are both connected to DDR4-2666 albeit with the ThunderX2 having 8 memory channels compared to 6 memory channels in the Skylake. The SG2042 has four memory controllers and only four memory channels, connected to DDR4-3200. Whilst there are fewer memory channels on the SG2042 than the other CPUs, it also has double the memory controllers than the Skylake and ThunderX2 CPUs, and also faster memory, but lags performance compared with those CPUs. Details around the memory subsystem on the SG2042 are difficult to come by, but it is our hypothesis that the memory controllers on the SG2042 are considerably less advanced than the other CPUs considered in this section.

The behaviour of the MG benchmark also helps explain one of the anomalies of the IS benchmark performance. It was our hypothesis that the L2 and L3 cache design was in part governing performance of the SG2042 compared to other CPUs, however the ThunderX2 also has the same cache design but was faster than the SG2042 for that benchmark. However, as seen for the MG benchmark, the SG2042 is also severely memory bandwidth bound and this likely explains the gap in performance between the SG2042 and ThunderX2 for the IS benchmark.

\subsection{Embarrassingly Parallel (EP)}

\begin{figure*}[htb]
    \centering
    {\includegraphics[width=0.8\textwidth]{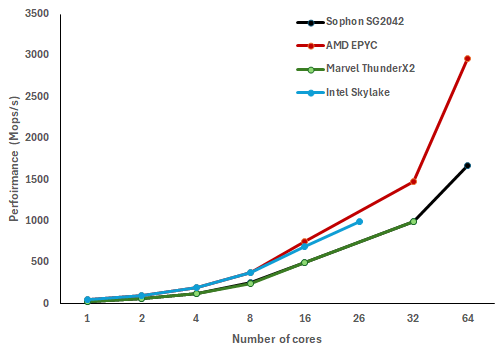}}
    \caption{EP benchmark performance (higher is better) parallelised via OpenMP}
    \label{fig:ep}
\end{figure*}

The Embarrassingly Parallel (EP) benchmark is compute bound, and results of this on our CPUs are illustrated in Figure \ref{fig:ep}. It can be seen that across the CPUs being tested, there are two groups; the SG2042 and ThunderX2 share very similar performance but with the SG2042's 64 cores then making a significant difference compared to the 32 cores of the ThunderX2. The EPYC and Skylake both deliver similar performance, which is greater than the SG2042 and ThunderX2, but the 26 cores of the Skylake are a disadvantage against the SG2042 which then significantly out performs the Skylake at 64 cores. The AMD EPYC performs best out of all the CPUs, especially at the larger core counts. 

This performance behaviour is in stark contrast to the IS and MG benchmarks, and demonstrates that the SG2042 and ThunderX2 deliver very similar compute performance at the same number of cores. This makes some sense given that they both provide 128-bit vectorisation, albeit with the ThunderX2 having two FPUs per core compared to one on the SG2042. By contrast, the Skylake and AMD EPYC CPUs provides wider vectorisation, 512-bit and 256-bit respectively and this in part helps explain the performance difference between these two groups. However, given the Skylake provides AVX512, and the AMD EPYC only AVX2, and that the ThunderX2 has two FPUs per core and the SG2042 only one, clearly the GCC compiler is not able to fully vectorise the code of this benchmark and make full use of the FPUs. 

Given that, at the largest number of cores each technology provides, the SG2042 performs second best for this benchmark out of all CPUs in our comparison, this demonstrates that for compute bound problems the large core count of the SG2042 is beneficial.

\subsection{Conjugate Gradient (CG)}

\begin{figure*}[htb]
    \centering
    {\includegraphics[width=0.8\textwidth]{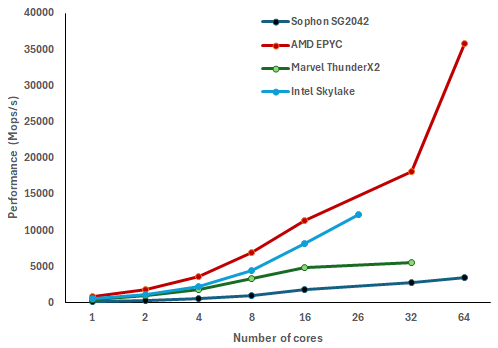}}
    \caption{CG benchmark performance (higher is better) parallelised via OpenMP}
    \label{fig:cg}
\end{figure*}

As can be seen from Table \ref{tab:npb_description}, the Conjugate Gradient (CG) benchmark also spends considerable time stalled on cache and DDR memory accesses, and this is because it comprises of irregular memory access and nearest neighbour communications. Figure \ref{fig:cg} illustrates the performance of this benchmark kernel across our CPUs. Given the performance of the IS and MG benchmarks it is no surprise that the SG2042 falls short of the other technologies, but it is closer to the ThunderX2 than we had expected delivering around 50\% the performance of the ThunderX2 at 32 cores. 

Potentially, what is making the difference here is in the size of the L3 cache, where the AMD EPYC has 16MB L3 shared between four cores (4MB per core), whereas the Skylake has 1.3MB L3 cache per core shared across all cores (potentially helped by the larger 1MB L2 cache). By contrast, both the ThunderX2 and SG2042 have the same size of 256MB L2 and 1MB L3 cache per core. This would help explain the performance differences, with the additional memory bandwidth limitations of the SG2042 causing additional the overhead which reduces performance further.

\subsection{Fast Fourier Transform (FT)}

\begin{figure*}[htb]
    \centering
    {\includegraphics[width=0.8\textwidth]{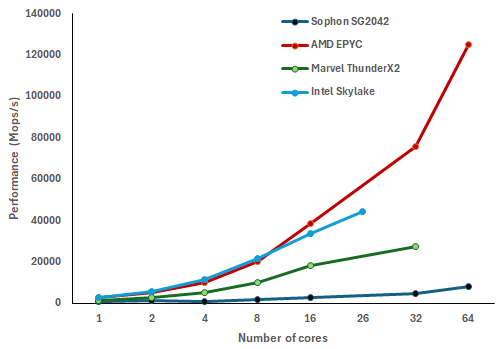}}
    \caption{FT benchmark performance (higher is better) parallelised via OpenMP}
    \label{fig:ft}
\end{figure*}

The fast Fourier Transform (FT) benchmark requires all-to-all communication between ranks, and the performance of this benchmark can be seen in Figure \ref{fig:ft}. As described in Table \ref{tab:npb_description}, for this benchmark on the Skylake there was some stalling due to cache and DDR access (13\% and 9\% respectively) but also for 18\% of the time DDR was under high utilisation. Once again, the SG2042 is significantly slower than the other CPUs, with the ThunderX2 sitting around half way between the performance of the x86 CPUs and the SG2042 and this is likely for the same reasons explored for the CG benchmark.

\subsection{Pseudo Applications}

Table \ref{tab:pseudo_apps} reports performance for the BT, LU and SP benchmarks and this is expressed as how many times faster this CPU is than the SG2042. Given the findings that the SG2042 struggles to perform when there is increased pressure on the memory subsystem, and based upon the stall numbers reported in Table \ref{tab:npb_description}, it was our expectation that the SG2042 would perform best for the BT benchmark and worst for the SP benchmark with the LU benchmark in between. This is broadly the case based upon the figures in Table \ref{tab:pseudo_apps}, where each number reports the number of times faster than each of the other CPUs are for each pseudo application at a specific core count. It can be seen that three other CPUs significantly outperform the SG2042 for the three pseudo applications.

\begin{table}[htb]
    \centering
    \caption{For each pseudo application, the number of times faster a specific CPU is than the SG2042 at the given number of cores}
    \label{tab:pseudo_apps}
    \begin{tabular}{|c|ccc|ccc|ccc|}
    \hline
      \textbf{Number} & \multicolumn{3}{c|}{\textbf{BT benchmark}}  & \multicolumn{3}{c|}{\textbf{LU benchmark}} & \multicolumn{3}{c|}{\textbf{SP benchmark}}\\
      \textbf{cores} & EPYC & Skylake & ThunderX2 & EPYC & Skylake & ThunderX2 & EPYC & Skylake & ThunderX2\\ 
      \hline
      16 & 3.23 & 3.28 & 2.43 & 3.65 & 4.15 & 2.86 & 5.01 & 3.91 & 3.65\\
      26 & 3.57 & 2.97 & 2.69 & 3.20 & 3.16 & 2.62 & 6.25 & 3.48 & 3.57\\
      32 & 3.68 & - & 2.64 & 3.40 & - & 2.94 & 5.26 & - & 3.22\\
      64 & 4.19 & - & - & 2.95 & - & - & 4.22 & - & - \\
    \hline
    \end{tabular}
\end{table}

\section{MPI vs OpenMP on the Sophon SG2042}

In Section \ref{sec:otherarch}, the NPB benchmarks were all run using the official NAS OpenMP implementation. This is sensible given that execution is occurring within a single memory space, however a question is whether, for best performance, one should write their parallel code using OpenMP or MPI within a node. These two models are very different, where OpenMP follows a thread based approach and by default all threads share the same memory area and explicit marshalling and protection of shared memory is required. In contrast, when using MPI tasks run as independent processes which share no data and instead communicate with each other via explicit messages.

\begin{figure*}[htb]
    \centering
    {\includegraphics[width=\textwidth]{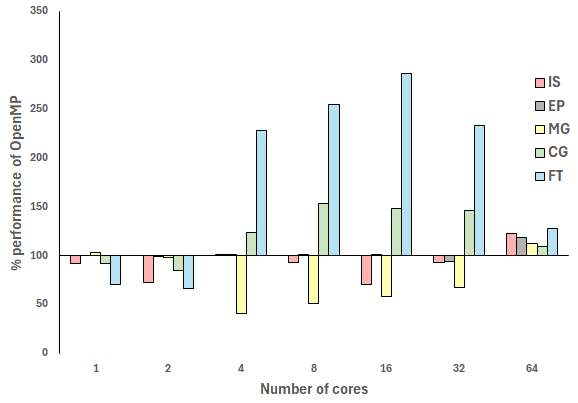}}
    \caption{Percentage performance of MPI based NPB benchmark compared to the OpenMP benchmark implementation}
    \label{fig:openmp_mpi}
\end{figure*}

As there are both OpenMP and MPI implementations of the NPB benchmark suite, we are able to undertake a direct comparison. This is illustrated in Figure \ref{fig:openmp_mpi} which reports the percentage performance delivered by the MPI implementation compared to the OpenMP version, for each benchmark at different numbers of cores. It can be seen that there are significant differences in performance between OpenMP and MPI for some configurations.

It can be observed, in Figure \ref{fig:openmp_mpi}, that over one or two cores the benchmarks do not benefit from MPI compared to OpenMP. However, from four cores upwards the CG and FT benchmarks run faster when using MPI than OpenMP. At 64 cores, all the MPI implementations run faster than their OpenMP counterparts. When exploring MPI against OpenMP on other architectures, MPI was always slower on the AMD EPYC and this was also true on the Skylake apart for the CG benchmark where MPI was on average two times faster than the OpenMP implementation. When profiling the CG MPI benchmark on the Skylake, it was found that clock ticks stalls due to cache reduced from 19\% in the OpenMP implementation to 5.5\% with the MPI version. The MPI implementation experienced no clock ticks stalled due to DDR accesses, down from 18\% for the OpenMP version. 

We surmise that the way these benchmarks are implemented means that the structure imposed by the MPI implementation tends to put less pressure on the memory subsystem when undertaking communications. For codes with heavy inter-core communication such as CG (nearest neighbour point to point) and FT (all to all collective) then this can be beneficial on the SG2042.

\section{Conclusions}

The Sophon SG2042 is an impressive RISC-V CPU and, using NASA's NAS Parallel Benchmark (NPB) suite, we have demonstrated that for benchmarks that closely represent ubiquitous HPC algorithms, especially in CFD, it significantly out performs existing RISC-V solutions. When compared against CPUs which implement other ISAs and whose use is widespread for high performance workloads, the Sophon SG2042 is outperformed by the AMD EPYC by between 1.77 and 15.06 times, the Intel Skylake between 0.59 and 5.98 times, and the Marvel ThunderX2 between 0.59 and 5.91 times. The SG2042 is most competitive for computationally bound codes, and its high core count ultimately out performed the Skylake and ThunderX2 for the EP benchmark. However the SG2042 CPU struggled with algorithms that are memory bandwidth or latency bound.

From this work we conclude that the memory subsystem of the SG2042 is a bottleneck, and Sophon recently announced the SG2044 which is reported to have three times the DDR memory bandwidth \cite{sophgo-roadmap}, as well as implementing RVV v1.0. This has the potential to provide a very significant performance improvement over the SG2042 for many of the benchmarks explored in this paper, and the improved memory performance will likely ameliorate the bottlenecks we have observed with the high core count continuing to deliver good computational performance. It is therefore our conclusion that the Sophon SG family of RISC-V CPUs has strong potential in HPC, and whilst the current generation SG2042 is an impressive first generation, based on announcements made by Sophon it looks likely that the SG2044 will address the key performance challenges that we have observed in this paper and make for a compelling RISC-V product family.

\section{Acknowledgements}
This work has been funded by the ExCALIBUR H\&ES RISC-V testbed. This work used the ARCHER2 UK National Supercomputing Service (https://www.archer2.ac.uk). The Fulhame HPE Apollo 70 system is supplied to EPCC as part of the Catalyst UK programme. For the purpose of open access, the author has applied a Creative Commons Attribution (CC BY) licence to any Author Accepted Manuscript version arising from this submission.

%
%
%
\bibliographystyle{splncs04}
\bibliography{citations}
\end{document}